\documentclass[10pt]{article}

\usepackage{amsfonts,amssymb,amsmath}		
\usepackage{graphicx}			
\usepackage{subfigure}		
\usepackage{caption}			
\usepackage{indentfirst}	
\usepackage{authblk}	 		
\usepackage{color}		 		
\usepackage{multirow}     
\usepackage{hyperref}     
\usepackage{fancyhdr}

\graphicspath{{./figures/}}

\hypersetup{
  colorlinks   = true, 
  urlcolor     = blue, 
  linkcolor    = blue, 
  citecolor    = red 	 
}

\usepackage[left=2.5cm,right=2.5cm,top=2.9cm,bottom=2.9cm,a4paper]{geometry}

\title{\bf Instability of Charged Anti-de Sitter Black Holes}
\author[1]{Bogeun Gwak\thanks{rasenis@sogang.ac.kr}}
\author[2]{Bum-Hoon Lee\thanks{bhl@sogang.ac.kr}}
\author[3]{Daeho Ro\thanks{dhro@sogang.ac.kr}}
\affil[1,2]{\small \it Center for Quantum Spacetime, Sogang University, Seoul 04107, Korea}
\affil[2]{\small \it Asia Pacific Center for Theoretical Physics, Pohang 37673, Korea}
\affil[2,3]{\small \it Department of Physics, Sogang University, Seoul 04107, Korea}

\date{}
\begin{document}
\maketitle
\thispagestyle{fancy}
\rhead{APCTP Pre2015-024}
\cfoot{}

\begin{abstract}
We study the instability of charged anti-de Sitter black holes in four or higher-dimension under fragmentation. The instability of fragmentation breaks the black hole into two black holes. We have found that the region near extremal or massive black holes become unstable under fragmentation. These regions depend not only on the mass and charge of initial black hole but also those of the fragmented one. The instability in higher-dimension is qualitatively similar to that of four-dimension. The detailed instabilities are numerically investigated.
\end{abstract}

\newpage
\setcounter{page}{1}

\section{Introduction} \label{sec:1}
A $D$-dimensional gravity theory in a bulk anti-de Sitter spacetime (AdS$_D$) corresponds to a $(D-1)$-dimensional conformal field theory (CFT$_{D-1}$) on the AdS$_D$ boundary. The correspondence is well known as the AdS/CFT duality \cite{Maldacena:1997re,Gubser:1998bc,Witten:1998qj,Aharony:1999ti}. Through this duality, the thermodynamic properties of a bulk gravity theory appear as those of the CFT on the boundary \cite{Witten:1998zw}. For example, AdS black holes can be described as finite temperature CFTs having Hawking temperature $T_H$, and the instabilities of AdS black holes are related to the phase transition of CFT systems. The dual CFT residing on a Reissner-Nordstr{\"o}m-AdS (RN-AdS) black hole is a field theory with a chemical potential for the charge of the black hole \cite{Hawking:1998kw,Hawking:1999dp}. An RN-AdS black hole becomes unstable in the presence of a charged scalar field \cite{Gubser:2008px}, which is interpreted as a superconductor instability \cite{Hartnoll:2008vx,Hartnoll:2008kx} in the dual CFT. Through AdS/CFT correspondence, a rotating holographic superconductor is the dual of a Kerr-Newman-AdS black hole \cite{Sonner:2009fk}. The charged black hole instability has been used to understand the phase transition of a holographic superconductor \cite{Franco:2009if,Bai:2014poa}.

Since the discovery of Higgs particle at the LHC\cite{ATLAS:2012ae,Chatrchyan:2012tx}, there have been detail studies on the Higgs potential in the high energy region, which suggests the present universe to be metastable. Then, the present false vacuum can tunnel into the more stable vacua, whose the decay and lifetime are calculated with the method of Coleman and De Luccia\cite{Coleman:1977py,Callan:1977pt,Coleman:1980aw}. Its lifetime turns out to be large enough compared with the age of universe. However, the gravitational impurities, e.g. black holes, generate the inhomogeneities and act as sites for vacuum decay. The inhomogeneities reduce an energy barrior to vaccum decay and also can significantly reduce the lifetime of the metastable state to millions of Planck times\cite{Burda:2015yfa,Burda:2015isa}. In this analysis, understanding of black holes including AdS one is important.

The stabilities of black holes have been investigated under perturbation or thermodynamics. A Schwarzschild black hole is stable under perturbation. An RN black hole is stable under the perturbation by the neutral \cite{Moncrief:1974gw,Moncrief:1974ng} and charged \cite{Hod:2013eea,Hod:2015hza} scalar fields. In AdS spacetime, a Schwarzschild-AdS black hole is perturbatively stable. The stability of an RN-AdS black hole depends on the theories. An RN-AdS black hole in Einstein-Maxwell gravity is stable under perturbation \cite{Konoplya:2008rq}. In the $\mathcal{N}=8$ gauged supergravity theory, the instability of an RN-AdS black hole comes from the tachyon mode of the scalar field \cite{Gubser:2000ec,Gubser:2000mm}. Also, the horizon of rotating black holes is stable under perturbation which is constrained by the particle energy equation \cite{Penrose:1969pc,Bardeen:1970zz,Wald1974548,Gwak:2011rp,Gwak:2012hq}. In contrast, thermodynamic stabilities are different from those of perturbation. A Schwarzschild black hole has negative heat capacity, so it is thermodynamically unstable. An RN black hole is also thermodynamically unstable within specific parameter regions \cite{Davies:1978mf,Anderson:1995fw,Pavon:1991kh}. A Schwarzschild-AdS black hole is unstable under thermodynamics. An RN-AdS black hole occurs during the second order phase transition \cite{Chamblin:1999hg}, so it is also unstable. The higher-dimensional rotating black hole, e.g. the Myers-Perry (MP) black hole, with large angular momentum can be unstable under fragmentation which breaks the black hole into several pieces \cite{Emparan:2003sy}. This provide dynamically the angular momentum upper bound for stability. Instability under fragmentation is a type of thermodynamic instability based on the black hole entropy. The instability of fragmentation occurs on MP-AdS black holes in perturbatively stable regions \cite{Gwak:2014xra}. According to dilaton-Gauss-Bonnet (DGB) theory, a static black hole becomes unstable under fragmentation in specific parameter regions \cite{Ahn:2014fwa}.

In this paper, we will investigate the instability of four or higher-dimensional charged AdS black holes under fragmentation into two black holes. Through the fragmentation instability of the charged AdS black hole, we can provide an entire prospect of the instability of black holes in Einstein gravity. The instability appears in the region near extremal charge or with large mass. It depends on not only initial black hole mass and charge but also the fragmentation ratio, i.e. the mass and charge ratio of final with initial black hole. The electric charge repulsion of black holes may cause the fragmentation instability.

This paper is organized as follows. In section \ref{sec:2}, a higher-dimensional charged AdS black hole is introduced. In section \ref{sec:3.1}, the fragmentation instability is represented and shown in several approximations. In section \ref{sec:3.2}, the fragmentation instabilities are numerically given as the phase diagrams with respect to black hole mass $M$ and charge $Q$. In section \ref{sec:3.3}, the phase diagrams are shown in terms of fragmentation ratios, $\epsilon_m$ and $\epsilon_q$ of the black hole with given mass $M$ and charge $Q$. In section \ref{sec:4}, we summarize our results.

\section{Charged Anti-de Sitter Black Holes} \label{sec:2}
The charged anti-de Sitter (AdS) black hole is a static solution of Einstein-Maxwell gravity with a negative cosmological constant. The charged AdS black hole with a mass $M$ and electric charge $Q$ is given in the $D$-dimensional spacetime \cite{Chamblin:1999hg,Chamblin:1999tk} as
\begin{equation} \label{eq:metric}
ds^2 \ =\ -f(r) dt^2 + \dfrac{1}{f(r)}dr^2 + r^2 d\Omega_{D-2},
\hspace{30pt} 
f(r) \ =\ 1 - \dfrac{2M}{r^{D-3}} + \dfrac{Q^2}{r^{2D-6}} + \dfrac{r^2}{\ell^2},
\end{equation} 
where $\Omega_{D-2}$ is $(D-2)$-dimensional sphere. We set the $c=G=1$ in this paper. The radius of horizon $r_h$  is a solution of the equation $f(r_h)=0$. It satisfies
\begin{equation} \label{eq:fr.re}
\dfrac{r_h^{2D-4}}{\ell^2} + r_h^{2D-6} - 2M r_h^{D-3} + Q^2 \ =\ 0.
\end{equation}

The radius of AdS curvature $\ell$ is described in terms of cosmological constant $\Lambda=-\frac{(D-1)(D-2)}{2\ell^2}$.  The asymptotic region is AdS spacetime of the AdS radius $\ell$. The black hole solution becomes Schwarzschild-AdS black hole for $Q=0$ and Reissner-Nordstrom black hole for $\ell\rightarrow \infty$. Hawking temperature $T_H$ and electric potential $\Phi_H$ are given at the black hole outer horizon $r_h$ 
\begin{equation} \label{eq:temp}
T_H \ =\ \dfrac{(D-3)\hbar}{4\pi}\left[\dfrac{r_h^{2D-6}-Q^2}{r_h^{2D-5}}+\dfrac{(D-1)r_h}{\ell^2}\right],
\hspace{30pt}
\Phi_H \ =\ \dfrac{Q}{r_h^{D-3}}.
\end{equation}
Bekenstein-Hawking entropy $S_{BH}$ is proportional to the horizon area $\mathcal{A}_H$
\begin{equation} \label{eq:entropy}
S_{BH} \ =\ \dfrac{\mathcal{A}_H}{4\hbar}=\dfrac{\Omega_{D-2} r_h^{D-2}}{4\hbar}.
\end{equation}
The black hole entropy also satisfies the 2nd law of thermodynamics, and the natural process of the black hole increases the entropy.

The function $f(r)$ has no horizon for given mass $M$ with large electric charge $Q$, because the power of mass term is smaller than charge term in $D\geq 4$. For arbitrary dimensions, the charged AdS black hole always has a bound value $Q$ for a given mass $M$ as an extremal black hole. The radius of horizon for an extremal black hole is a solution of
\begin{equation} \label{eq:fr.ext}
\dfrac{D-2}{D-3} \dfrac{r_h^{D-1}}{\ell^2} + r_h^{D-3} - M \ =\ 0.
\end{equation}
We may define the dimensionless variables by rescaling for the AdS radius
\begin{equation} \label{eq:dimensionless}
\dfrac{s}{\ell} \rightarrow s, \qquad \dfrac{t}{\ell} \rightarrow t, \qquad \dfrac{r}{\ell} \rightarrow r, \qquad \dfrac{M}{\ell^{D-3}} \rightarrow M, \quad \text{and} \quad \dfrac{Q}{\ell^{D-3}} \rightarrow Q.
\end{equation}
This is equivalent to putting $\ell=1$ in the above expressions.

\section{Instability from Fragmentation} \label{sec:3}
Let us assume that the charged AdS black hole is broken into two black holes and moved to far from each other, slowly. We treat one of the fragmented black holes as a test particle moving in a charged AdS black hole spacetime given by the other one. The test particle with a mass $M$ and charge $Q$ has the energy $E$ given by
\begin{equation} \label{eq:geodesic}
E^2 \ =\ M^2 \left(1 - \dfrac{2M}{R^{D-3}} + \dfrac{Q^2}{R^{2D-6}} + \dfrac{R^2}{\ell^2}\right),
\end{equation}
where $R$ is the impact parameter of a black hole. Since the black holes are located far from each other, we can approximate $f(R) \cong 1 + \frac{R^2}{\ell^2}$. Furthermore, if we suppose the radius of AdS curvature is sufficiently larger than the impact parameter, $\ell \gg R$, then the gravitational interaction of the black hole and effect of the cosmological constant are negligible and so equation \eqref{eq:geodesic} is reduced to 
\begin{equation}
E^2 \ \cong\ M^2.
\end{equation}
In this set up, the total mass and charge are preserved in the initial and final states. 

The black hole prefers larger entropy state. Thus, the fragmentation occurs when the entropy of initial black hole state is smaller than that of final black hole state. The entropy of a black hole is given by equation \eqref{eq:entropy} which is proportional to the power of radius at the horizon of a black hole. Let us consider the initial black hole with a mass $M$ and charge $Q$ is fragmented into two black holes. The mass and charge are divided into two parts under the mass and charge conservation. The mass and charge ratio are defined as $0 \leq \epsilon_m \leq 1$ and $0 \leq \epsilon_q \leq 1$. After fragmentation, one has the mass $\epsilon_m M$ and charge $\epsilon_q Q$ and the other has the mass $(1 - \epsilon_m) M$ and charge $(1 - \epsilon_q) Q$. Then, we may introduce the entropy ratio for the initial and final black hole states 
\begin{equation} \label{eq:ratio}
R \ =\ \dfrac{S_{BH,f}}{S_{BH,i}} \ =\ \dfrac{r_h \big[ \epsilon_m M, \epsilon_q Q \big]^{D-2} + r_h \big[ (1-\epsilon_m) M, (1-\epsilon_q) Q \big]^{D-2}}{r_h [ M, Q ]^{D-2}}.
\end{equation}
where $r_h [M,Q]$ is the radius of black hole horizon with a given mass $M$ and charge $Q$. The black hole is unstable and fragmented into two black holes when the entropy ratio is larger than $1$ and the other case is stable.

\subsection{Analytical Approximation for the Instability} \label{sec:3.1}
Before we check the instability of charged AdS black holes numerically, the instability is obtained by analytically with both small and large parameter limit in the general dimensions $D \geq 4$. For simplicity, we assume that the black hole is fragmented into two identical black holes for the analytic calculation, i.e. the fragmentation ratio of mass $\epsilon_m$ and charge $\epsilon_q$ satisfies $\epsilon_m=\epsilon_q=1/2$. Then, the entropy ratio in equation \eqref{eq:ratio} is simplified as
\begin{equation} \label{eq:ratio_identical}
R \ =\ 2 \left(\dfrac{r_h \big[ M/2, Q/2 \big]}{r_h [ M, Q ]} \right)^{D-2}.
\end{equation}

Let us consider the instability of charged AdS black holes with small mass and charge limit, $M,Q \ll 1$. Note that $Q \leq M$ due to the constraint. Then, the extremal black hole is enough to take only a small mass limit. In this limit, the entropy ratio can be rewritten
\begin{equation}
R \ =\ 2^{-\tfrac{1}{D-3}} < 1.
\end{equation}
Since the entropy ratio is always smaller than $1$, the charged AdS black hole with small mass and charge limit is stable under fragmentation. 

Now, consider the large mass and small charge limit, $M \gg 1$ and $Q \ll 1$. Then, the charge term in equation \eqref{eq:fr.re} is approximately zero. In addition, $r_h^{2D-4} \gg r_h^{2D-6}$ with every $D$ for the large mass limit. It leads to the entropy ratio
\begin{equation}
R \ =\ 2^{\tfrac{1}{D-1}} > 1,
\end{equation}
and it is always larger than 1. Thus, the charged AdS black hole with large mass and small charge limit is unstable under fragmentation.

Finally, we consider the large mass and charge limit. For simplicity, choose $Q=Q_{\text{ext}}$. In the leading order, the entropy ratio
\begin{equation}
R \ =\ 2^{\tfrac{1}{D-1}} > 1,
\end{equation}
where it is always larger than 1. Therefore, the extremally charged AdS black hole with large mass limit is unstable under fragmentation.

\begin{table}[ht]
\centering
\begin{tabular}{c|c|c}
& $M \ll 1$ & $M \gg 1$ \\ \hline 
$Q = Q_{\text{ext}}$ & \multirow{2}{*}{$ 2^{-\tfrac{1}{D-3}} \hspace{10pt}$ (stable)} & $2^{\tfrac{1}{D-1}} \hspace{10pt}$ (unstable) \\
$Q \ll 1$ & & $2^{\tfrac{1}{D-1}} \hspace{10pt}$ (unstable)
\end{tabular}
\caption{Instability of black holes under the limit of small and large parameter values with $\epsilon_m=\epsilon_q=1/2$.}
\label{tab:analytic}
\end{table}

The results are arranged in table \ref{tab:analytic}. The table shows the existence of both stable and unstable regions in $M$ and $Q$ parameter space. This suggests the existence of boundary. The exact boundary between stable and unstable region, which will be investigated numerically in the following section.

\subsection{Instability on \texorpdfstring{$M$-$Q$}{Mass-Charge} Diagram} \label{sec:3.2}
We now investigate the instability of charged AdS black holes numerically in the dimensionless mass $M$ and charge $Q$ parameter space. In this section, The decay with uniform ratio, $\epsilon = \epsilon_m = \epsilon_q$ is considered. The mass and charge ratio $Q/M$ of fragmented are the same as that of the original black hole. Other values of the ratios give qualitatively similar results which will be analyze in the next section. The entropy ratio in the uniform ratio is given
\begin{equation} \label{eq:ratio12}
R \ =\ \dfrac{S_{BH,f}}{S_{BH,i}} \ =\ \dfrac{r_h \big[ \epsilon M, \epsilon Q \big]^{D-2} + r_h \big[ (1-\epsilon) M, (1-\epsilon) Q \big]^{D-2}}{r_h [ M, Q ]^{D-2}}.
\end{equation}

\begin{figure}[ht]
\centering
\subfigure[(a)][The diagram for $D=4$]{\includegraphics[width=0.46\textwidth]{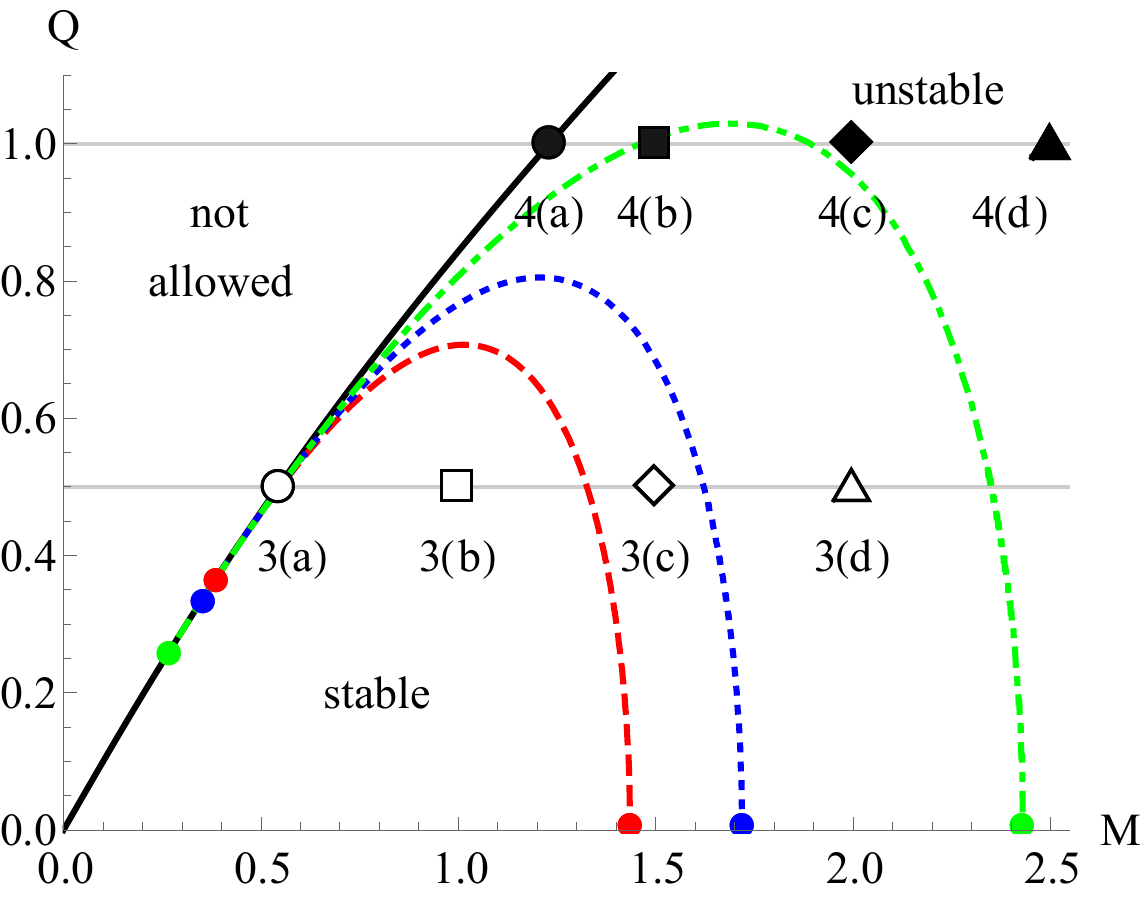}}
\hspace{0.05\textwidth}
\subfigure[(b)][The diagram for $D=6$]{\includegraphics[width=0.46\textwidth]{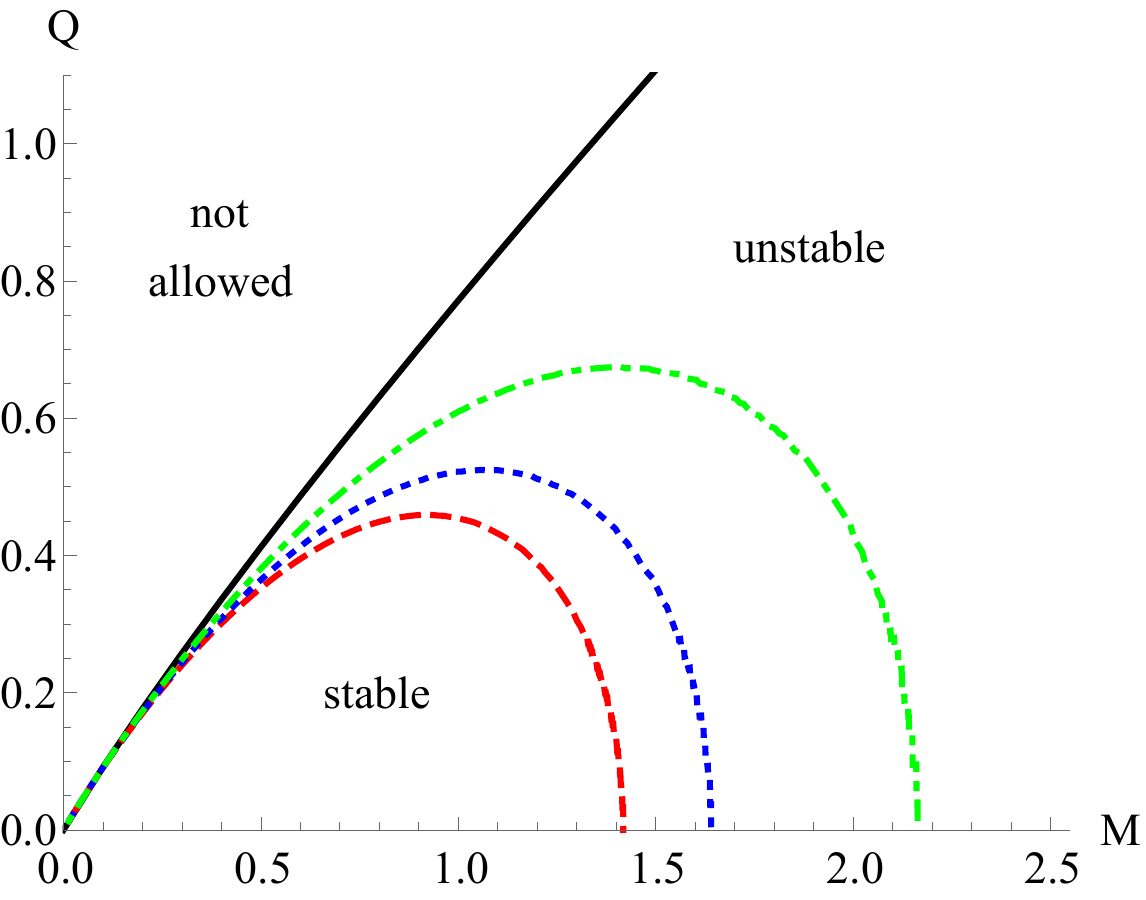}}
\caption{Phase diagrams for the instability of charged AdS black holes with respect to the mass and charge. Black line represents the extremal black hole. Dashed red, dotted blue, and dot-dashed green line represent the boundaries of black hole instability with fragmentation ratios $\epsilon=1/2$, $1/4$ and $1/8$, respectively. Markers indicate the parameter choices of figures \ref{fig:emq_diagram1} and \ref{fig:emq_diagram2} as we labelled.}
\label{fig:mq_diagram}
\end{figure}

The detailed phase diagram is summarized in figure \ref{fig:mq_diagram}. The extremal black hole bound is represented by the black solid line. The upper region of black solid line is not allowed because the initial black hole cannot have a charge larger than $Q_{\text{ext}}$. Below the black solid line, we plot three different instability boundaries which are represented by dashed red, dotted blue, and dot-dashed green lines corresponding to the fragmentation ratio $\epsilon=1/2$, $1/4$ and $1/8$, respectively. Colored dots represent the beginning and ending of them. For each curve, the stable region is inside of curve while outside is unstable. The plot for higher dimensions looks very similar to that for $D=4$, as can be seen in figure \ref{fig:mq_diagram} (b) for $D=6$.

The instability curve begins at the extremal black hole line with mass $M_{\text{ext},\epsilon}$ and ends at Schwarzschild-AdS black hole line with mass $M_{\text{max},\epsilon}$ which are denoted by colored dots in figure \ref{fig:mq_diagram} (a). Along the instability boundary for given $\epsilon$, $M_{\text{ext},\epsilon}$ and $M_{\text{crit},\epsilon}$ corresponds to the smallest and largest mass, respectively. All the extremal black holes with mass larger than $M_{\text{ext},\epsilon}$ will be unstable. Hence $M_{\text{ext},\epsilon}$ can also be interpreted as the instability boundary point for the extremal charged black holes. Similarly, $M_{\text{crit},\epsilon}$ can be interpreted as the instability boundary point for Schwarzschild-AdS black hole. On each boundary curve, there exist a maximum charge value $Q_{\text{max},\epsilon}$. The corresponding mass of that is denoted by $M_{\text{max},\epsilon}$. The curve grows up until $M_{\text{max},\epsilon}$ and goes down after words. The behavior of instability is divided by two regions and drastically changes depending whether the black hole has a mass smaller or larger than $M_{\text{max},\epsilon}$. For example with $\epsilon=1/2$, the instability curve begins at the extremal black hole line with mass $M_{\text{ext},1/2} = 0.388428$ and it grows up until $M_{\text{max},1/2} = 1.00843$ with charge $Q_{\text{max},1/2} = 0.706962$ After that, the curve ends down and ends up at Schwarzschild-AdS black hole with mass $M_{\text{crit},1/2} = 1.4355$. 

Let us consider the black hole with a mass larger than $M_{\text{max},\epsilon}$. For the region between two different boundaries specified by two different $\epsilon$ values, the black hole is unstable under the larger value of $\epsilon$ while stable under the smaller value of $\epsilon$. For example, suppose the parameters of initial black hole is located between red and blue boundaries. Then, the black hole is unstable under the $\epsilon=1/2$ represented by red boundary and stable under the $\epsilon=1/4$ represented by blue boundary. The stable region is decreasing and becomes minimal when $\epsilon=1/2$. The black hole will be more unstable as mass or charge increase because it has more channels to be fragmented through channels. Especially, Schwarzschild-AdS black hole with mass larger than $M_{\text{crit},1/2}$ is unstable. However, these black holes are unstable with $\epsilon$ larger than some finite value. In other words, it is non-perturbatively unstable while perturbatively stable.

\begin{figure}[ht]
\centering
\subfigure[(a)][Instability boundary of extremal AdS black hole]{\includegraphics[width=0.46\textwidth]{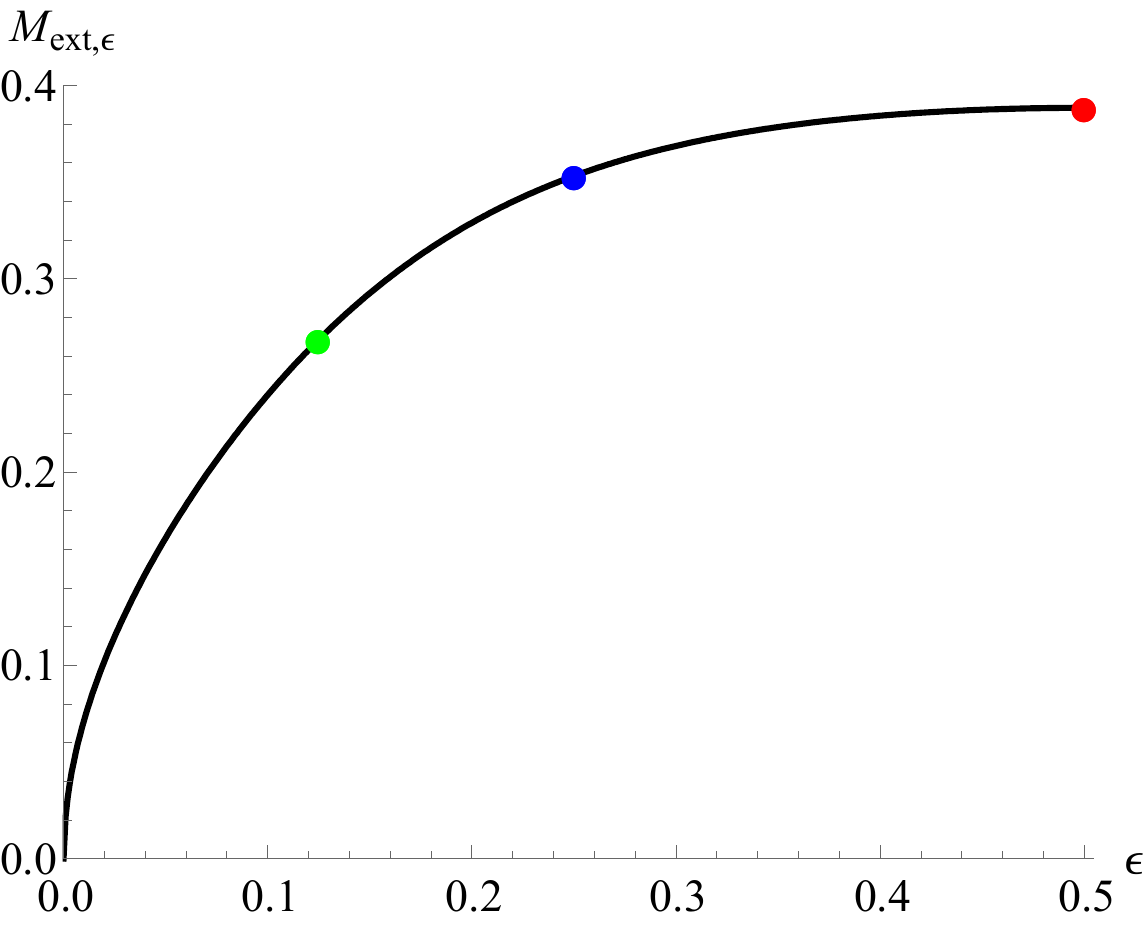}}
\hspace{0.05\textwidth}
\subfigure[(b)][Instability boundary of Schwarzschild-AdS black hole]{\includegraphics[width=0.46\textwidth]{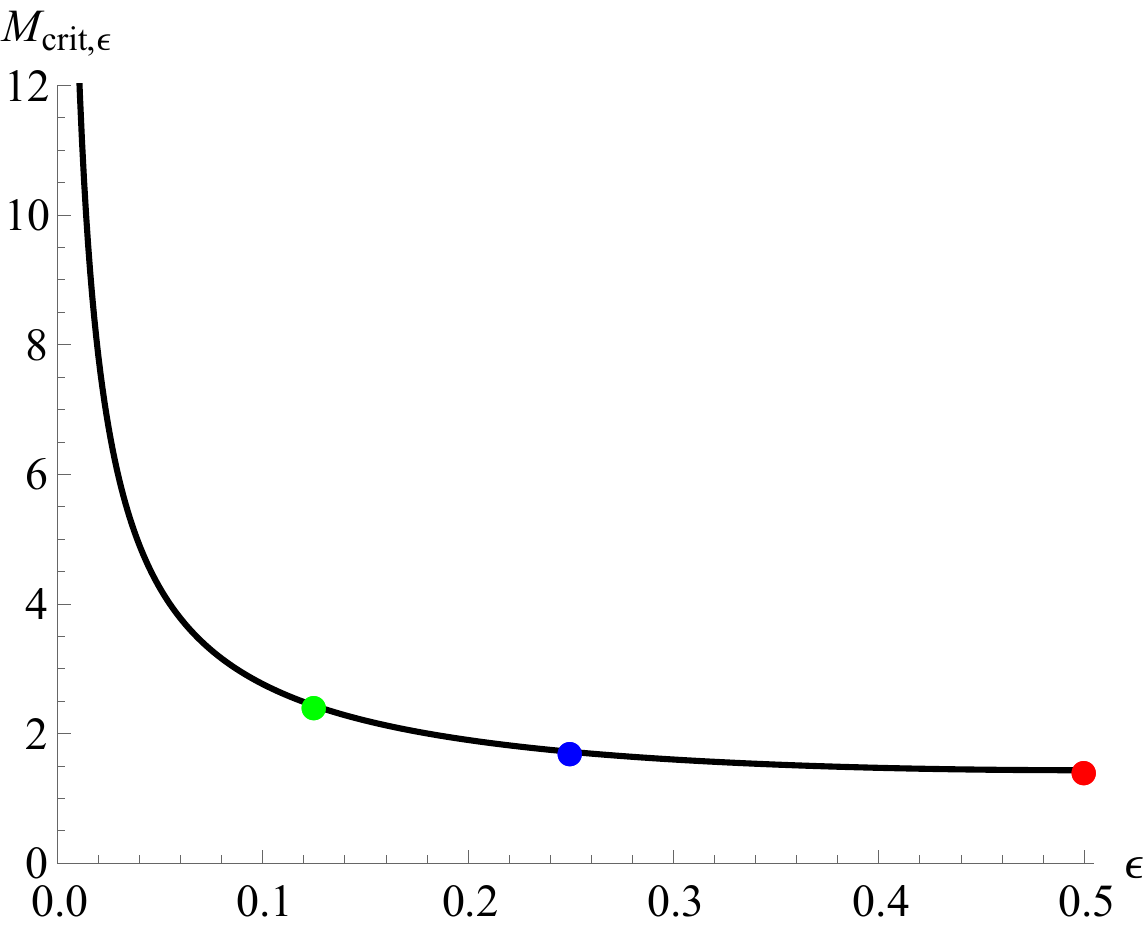}}
\caption{The plots of $M_{\text{ext},\epsilon}$ and $M_{\text{crit},\epsilon}$ with respect to fragmentation ratio $\epsilon$. We set $D=4$.}
\label{fig:mem_diagram}
\end{figure}

On the contrary, the behavior of black hole instability with a mass smaller than $M_{\text{max},\epsilon}$ is different. The black hole will now be more unstable as mass decreases or charge increases. In addition, the relation between stable region and the mass of black hole is not simple. For a given mass, the value $\epsilon$ gives a minimal stable region which is not necessarily $\epsilon=1/2$, but arbitrary $\epsilon$ which goes to zero as mass $M$ goes to zero while it increases toward $1/2$ and eventually become $1/2$ as mass $M$ increases. This phenomena occurs because the instability boundaries intersect each other, though it is hard to observe in figure \ref{fig:mq_diagram}. So, the plot of $M_{\text{ext},\epsilon}$ and $M_\epsilon$ in figure \ref{fig:mem_diagram} are introduced.

The shape of boundary is qualitatively similar for higher dimensions, $D \geq 4$. However, the stable region for a given $\epsilon$ shrinks as the dimension increases, which can be seen in figure \ref{fig:mq_diagram} (b). That means the possible range of charge for the stable black hole decreases with a given mass smaller than $M_{\text{crit},\epsilon}$. Furthermore, the value of $M_{\text{crit},\epsilon}$ also decreases. Thus, the black hole becomes unstable more easily in higher dimensions. 

To study the instability of black hole fragmentation with small mass in more detail, we plot $M_{\text{ext},\epsilon}$ and $M_{\text{crit},\epsilon}$ in figure \ref{fig:mem_diagram}. The colored dots represent $M_{\text{ext},\epsilon}$ and $M_{\text{crit},\epsilon}$ with respect to the selected values of $\epsilon$ which are used in figure \ref{fig:mq_diagram} (a). For smaller $\epsilon$, the instability boundary curve in figure \ref{fig:mq_diagram} starts from smaller $M_{\text{ext},\epsilon}$ and ends at larger $M_{\text{crit},\epsilon}$. On the other hands, for larger $\epsilon$, the instability boundary curve starts from larger $M_{\text{ext},\epsilon}$ and ends at smaller $M_{\text{crit},\epsilon}$. Therefore, any two curves with two different $\epsilon$ values must cross each other. We confirmed numerically that the crossing occurs very close to the extremal black hole line, and can be seen in figure \ref{fig:mq_diagram} only if we magnify significantly. Note that $M_{\text{crit},\epsilon}$ goes to infinity when $\epsilon$ goes to zero. This is consistent with the fact that Schwarzschild-AdS black hole is perturbatively stable.

Now, we consider the flat space limit by taking $\ell \rightarrow \infty$. The rescaled dimensionless parameters in figure \ref{fig:mq_diagram} in terms of the actual dimensionful quantities are $\frac{M}{\ell^{D-3}}$ and $\frac{Q}{\ell^{D-3}}$, respectively. Then, the limit $\ell \rightarrow \infty$ can be thought as the limit $M \rightarrow 0$ and $Q \rightarrow 0$. Thus, the flat space limit is equal to the small mass and small charge limit. This was shown by analytic method for $\epsilon=1/2$ as given in table \ref{tab:analytic} and by numerical method for arbitrary $\epsilon$ as shown in figure \ref{fig:mq_diagram}. Therefore, all the black holes are stable in the flat space limit under fragmentation with arbitrary uniform fragmentation ratio.

The instability of black hole fragmentation is determined by competition between gravitational attraction and effective repulsion which include the electric repulsion and repulsion from the AdS effect. In small ratio, the gravitational attraction is smaller than in the large ratio, but the repulsion from the AdS effect is uniformly contributed to the effective repulsion. Then, the fragmented black holes undergo larger effective repulsion in small ratio. This can explain why the extremal black hole is unstable for small ratio. However, the instability of large mass black hole appears in the first at $\epsilon=1/2$. In general, the electric repulsion is the largest when the fragmented black holes have the same charge. The gravitational attraction may be smaller than repulsion, because the center of two fragmented black hole have a distant to twice of horizon.

\subsection{Instability on \texorpdfstring{$\epsilon_m$-$\epsilon_q$}{Fragmentation Mass-Charge Ratio} Diagram} \label{sec:3.3}

\begin{figure}[ht!]
\centering
\subfigure[(a)][The diagram for $M=M_{\text{ext}}$]{\includegraphics[width=0.4\textwidth]{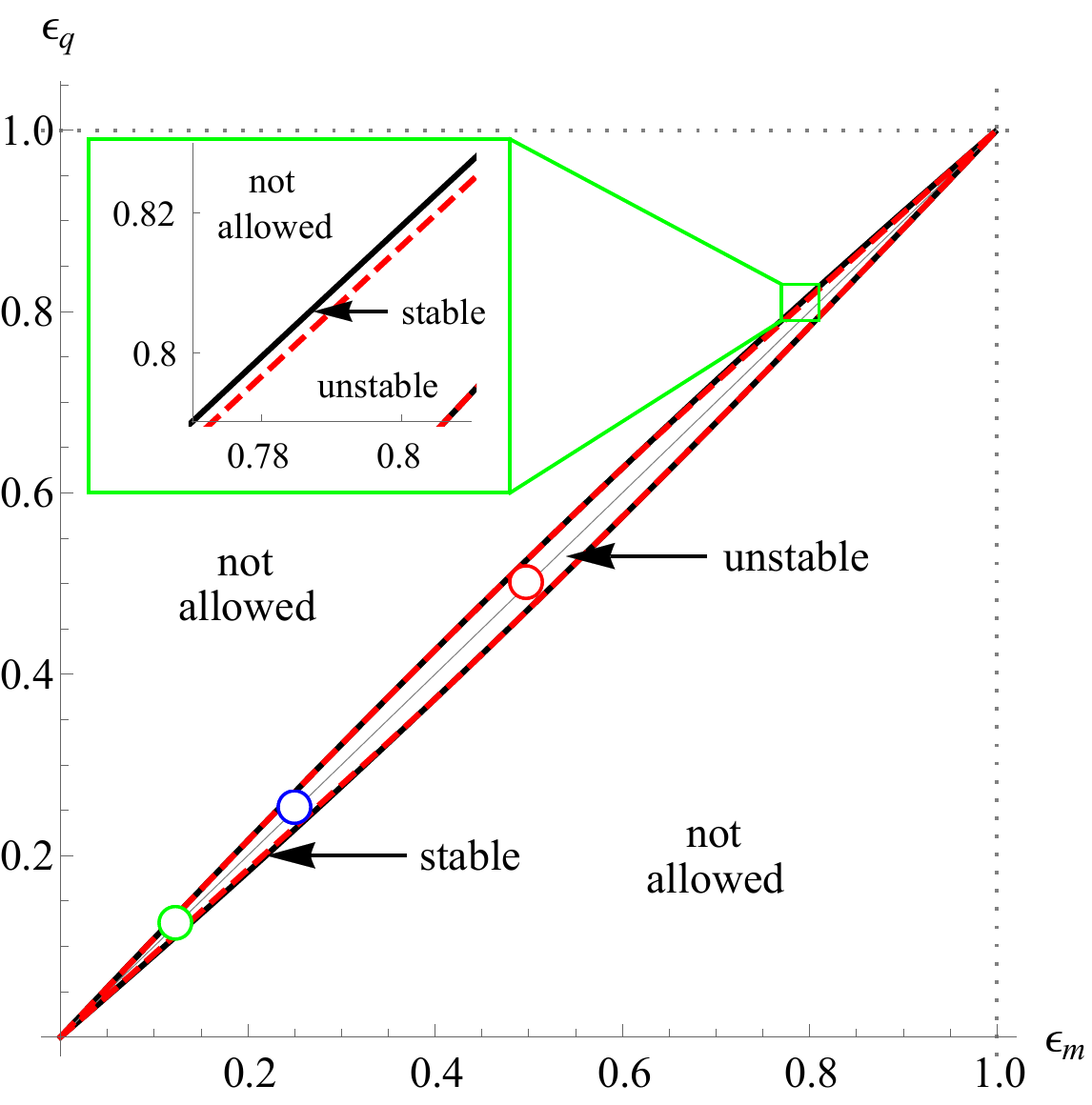}}
\hspace{0.05\textwidth}
\subfigure[(b)][The diagram for $M=1.0$]{\includegraphics[width=0.4\textwidth]{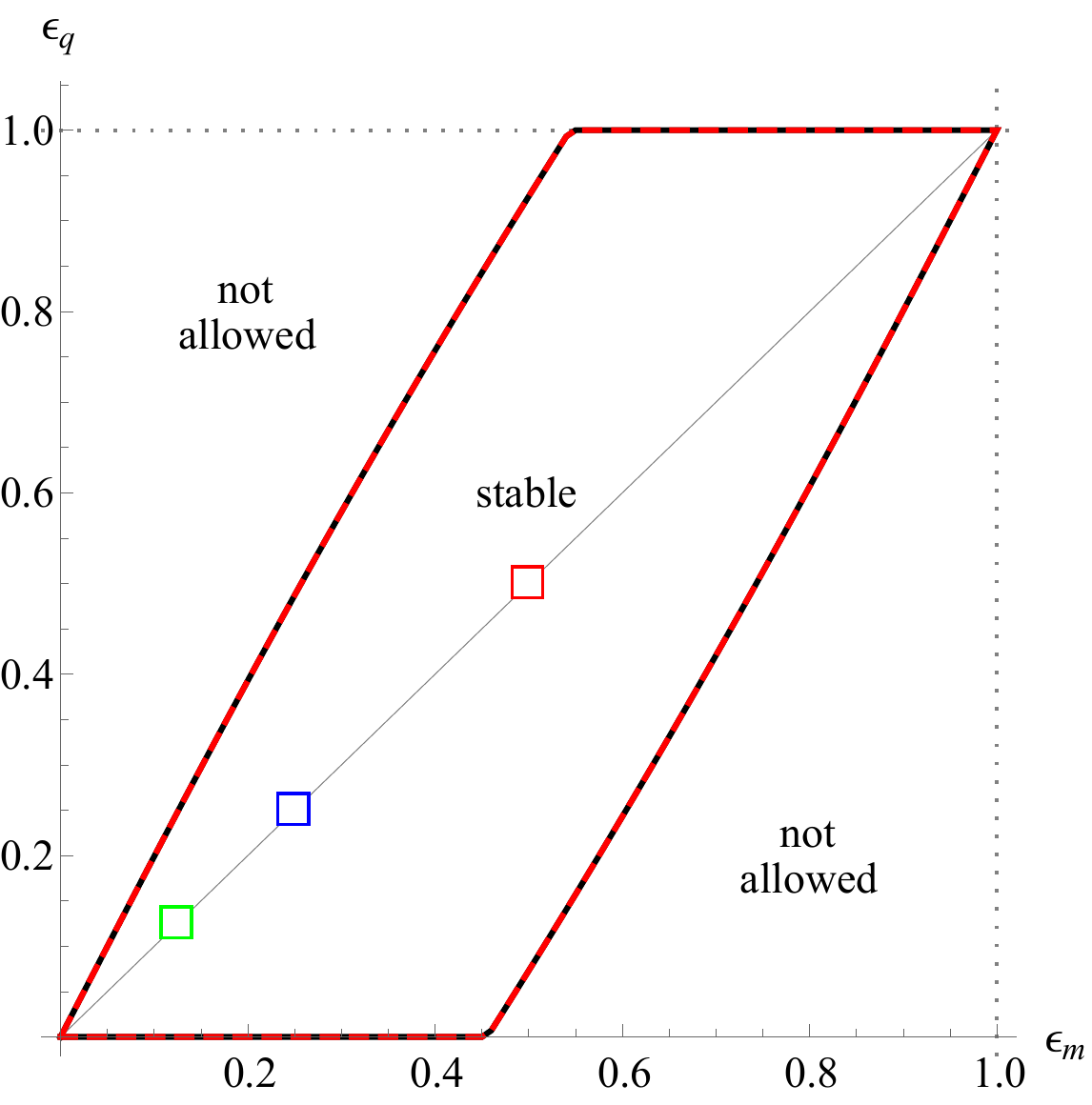}}
\\
\subfigure[(c)][The diagram for $M=1.5$]{\includegraphics[width=0.4\textwidth]{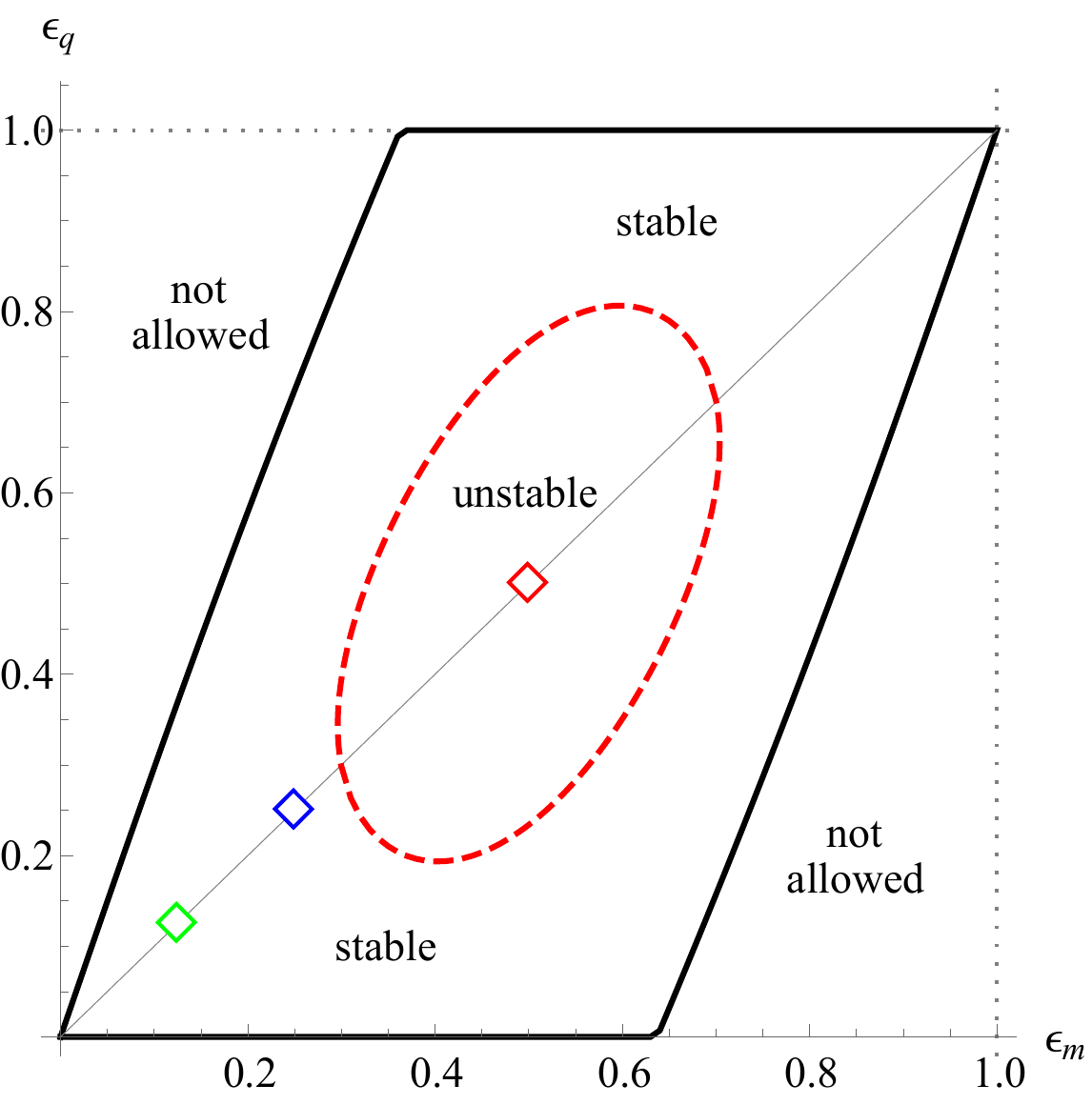}}
\hspace{0.05\textwidth}
\subfigure[(d)][The diagram for $M=2.0$]{\includegraphics[width=0.4\textwidth]{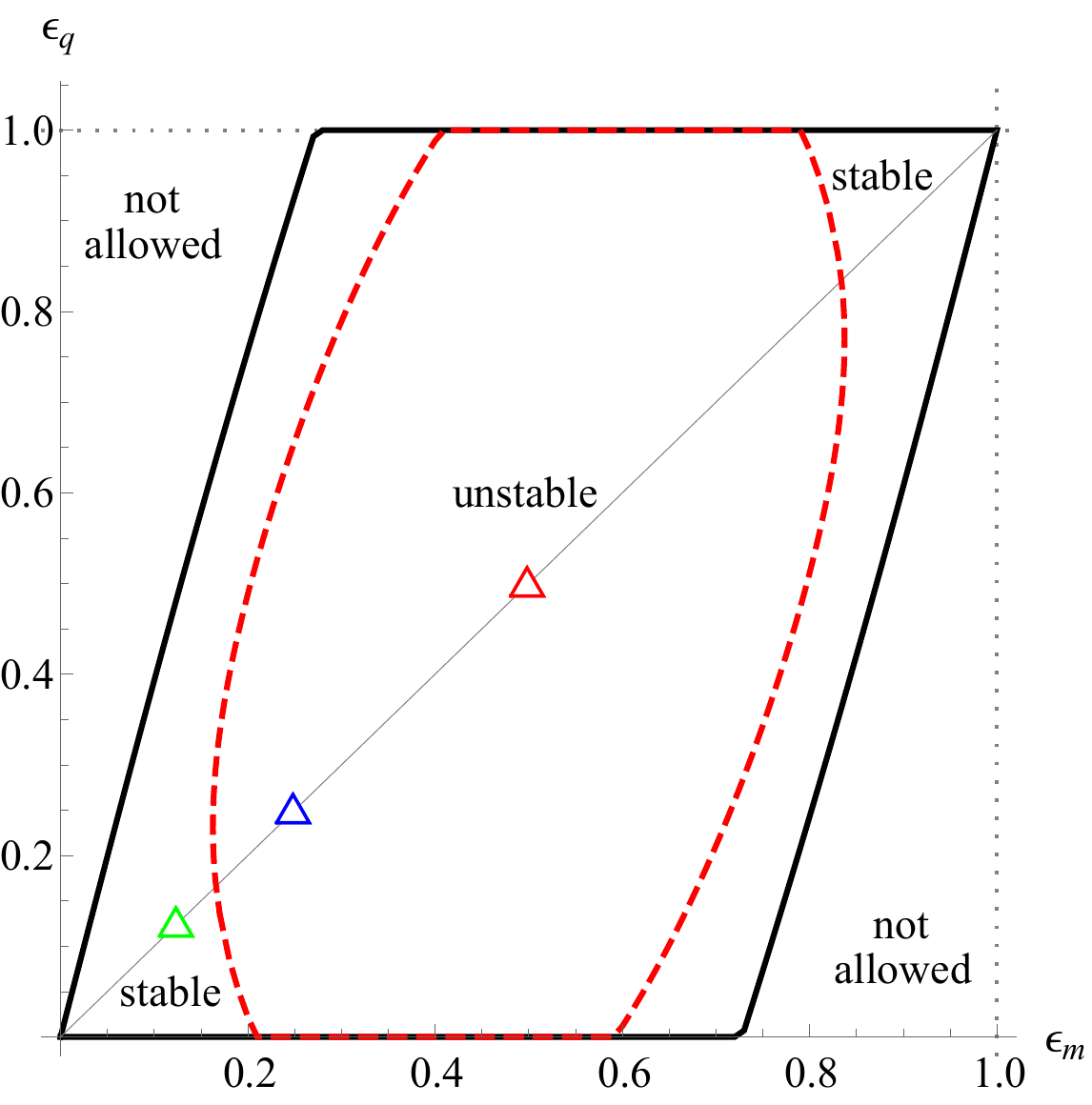}}
\caption{Phase diagrams for the instability of charged AdS black holes with respect to mass and charge fragmentation ratio, $\epsilon_m$ and $\epsilon_q$. In this figure, we set $D=4$ and $Q=0.5$. Black line represents the possible black hole fragmentation under the extremal condition. Dashed red line is the instability boundary of black hole fragmentation. Red, blue and green markers represent the parameter choices $\epsilon_m=\epsilon_q=1/2$, $1/4$, and $1/8$, respectively. $M_{\text{ext}}$ is a mass of extremal black hole with a given charge. Phase diagram for higher-dimensional black hole looks very similar to $D=4$ phase diagram.}
\label{fig:emq_diagram1}
\end{figure}

\begin{figure}[ht!]
\centering
\subfigure[(a)][The diagram for $M=M_{\text{ext}}$]{\includegraphics[width=0.4\textwidth]{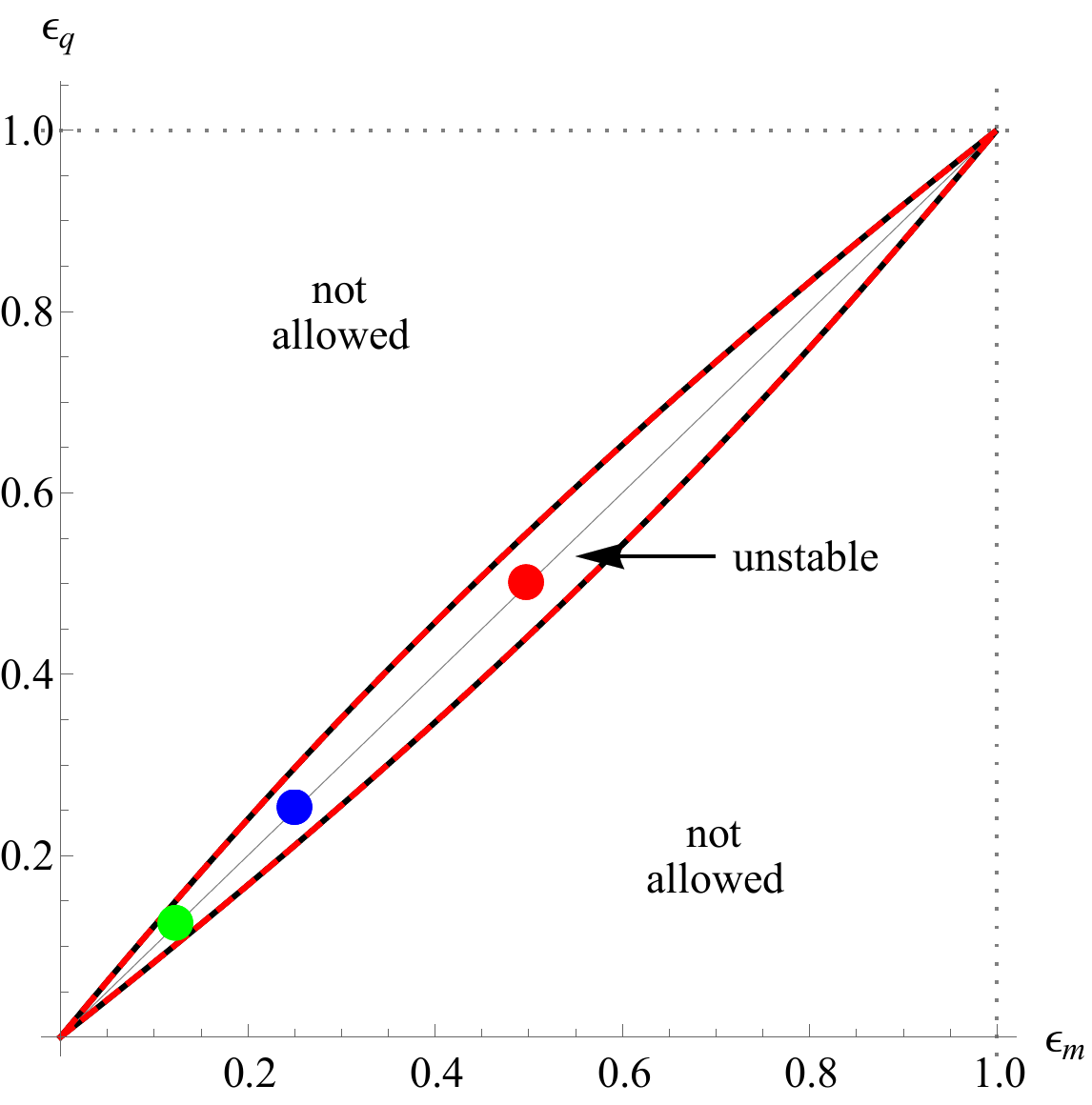}}
\hspace{0.05\textwidth}
\subfigure[(b)][The diagram for $M=1.5$]{\includegraphics[width=0.4\textwidth]{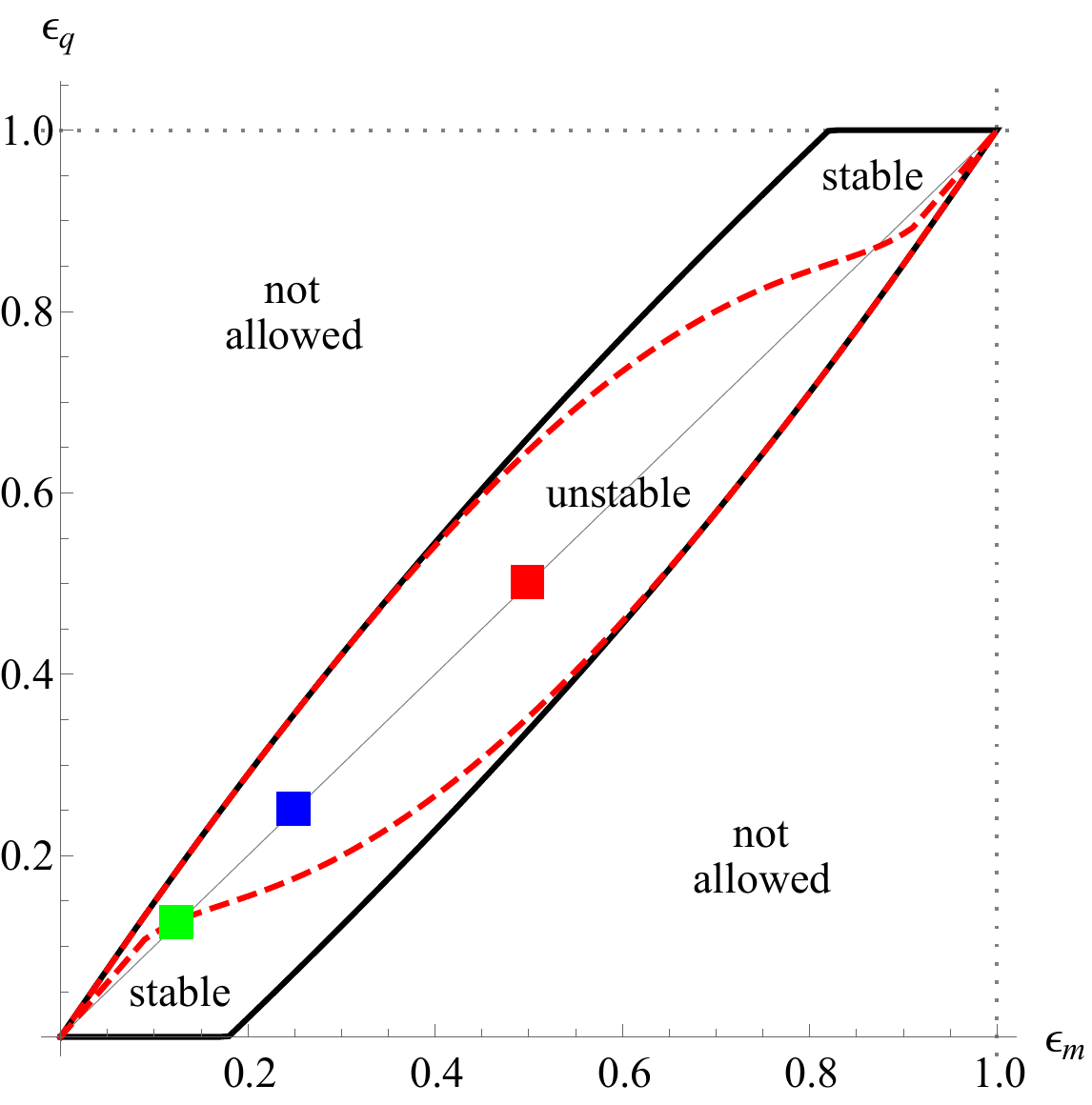}}
\\
\subfigure[(c)][The diagram for $M=2.0$]{\includegraphics[width=0.4\textwidth]{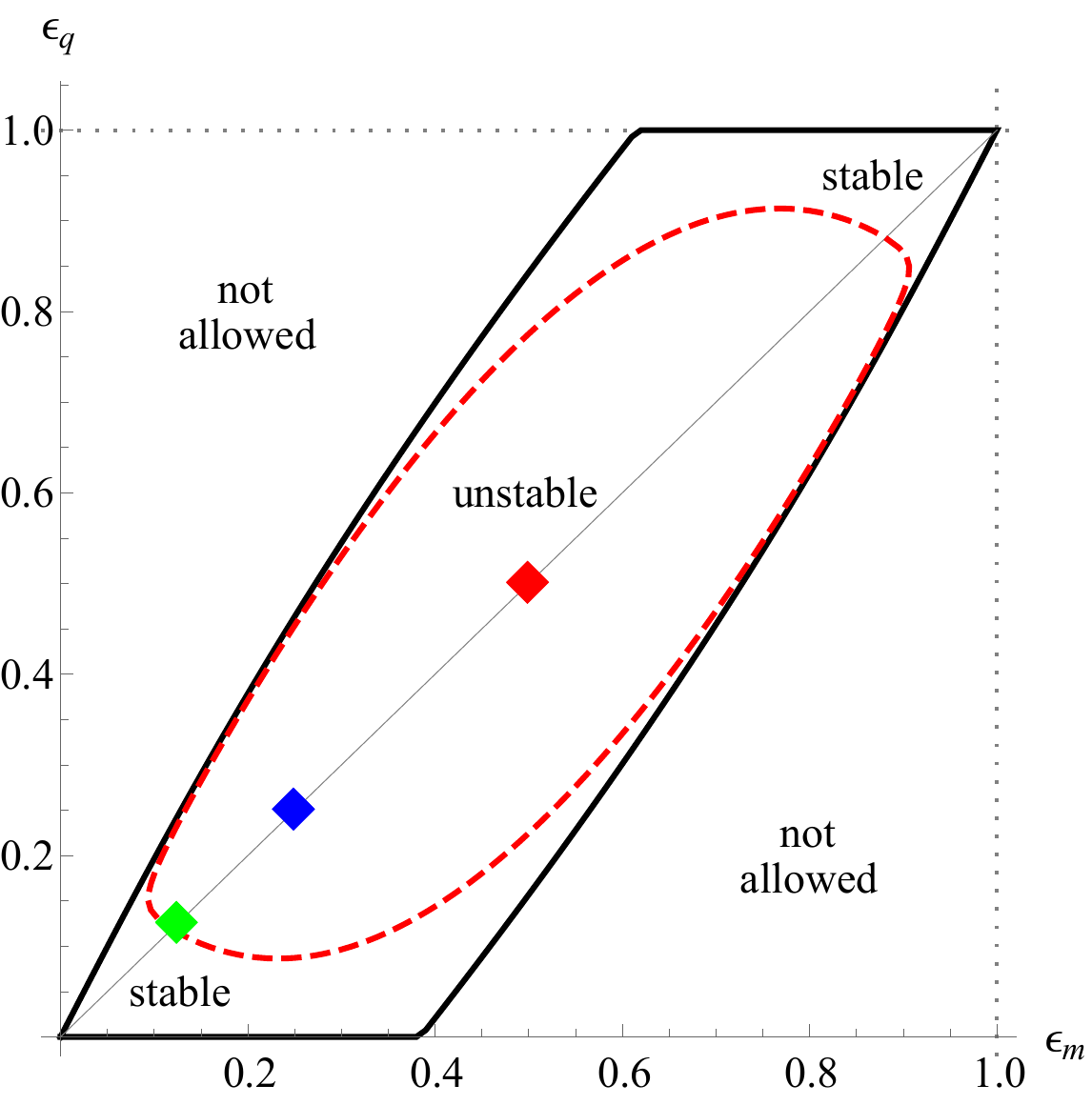}}
\hspace{0.05\textwidth}
\subfigure[(d)][The diagram for $M=2.5$]{\includegraphics[width=0.4\textwidth]{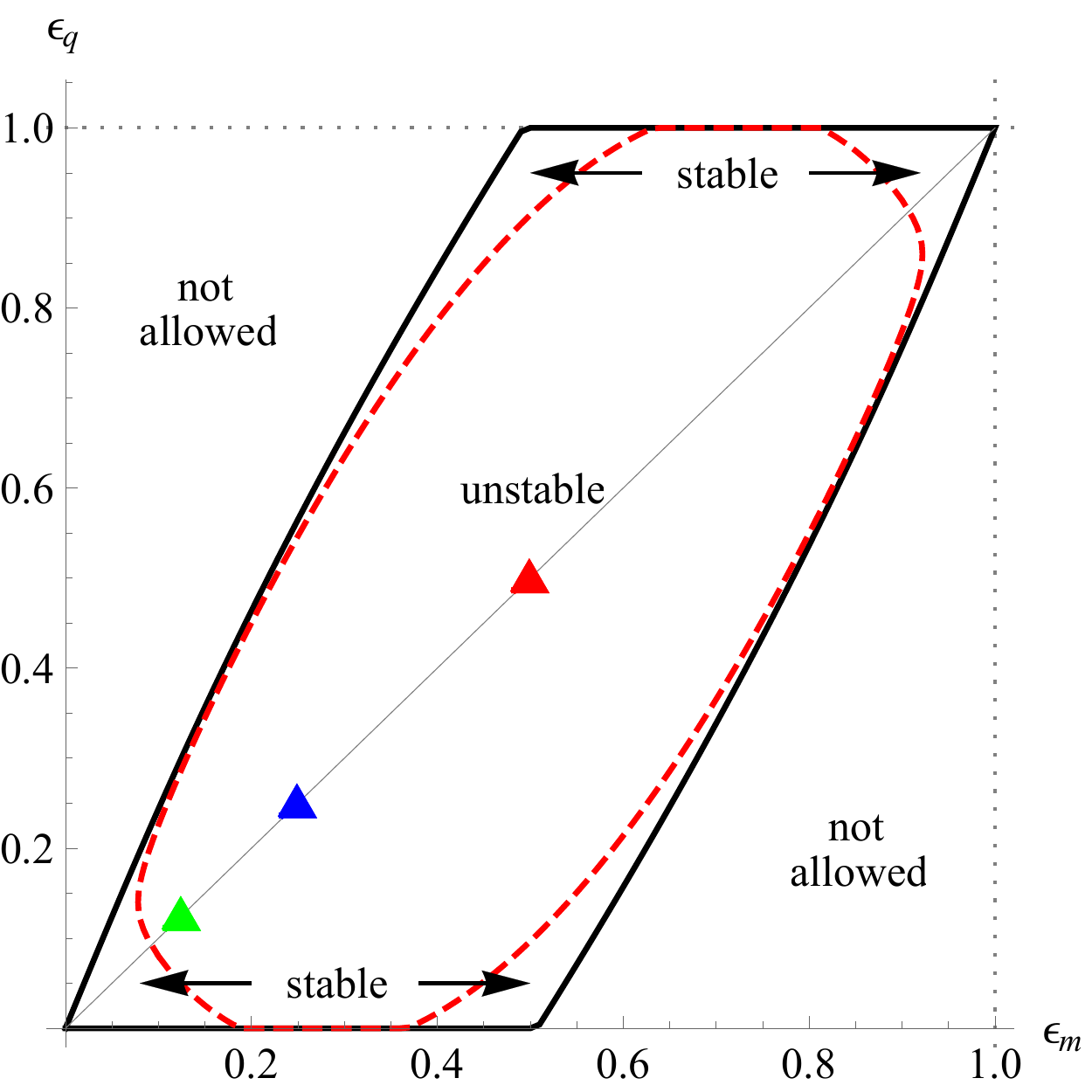}}
\caption{Phase diagrams for the instability of charged AdS black holes same as figure \ref{fig:emq_diagram1}. In this figure, we set $D=4$ and $Q=1.0$.}
\label{fig:emq_diagram2}
\end{figure}

We discuss the instability of charged AdS black holes under fragmentation into two black holes with arbitrary fragmentation ratio $\epsilon_m $ and $\epsilon_q $, numerically. The instability of black holes depends on the fragmentation ratio of mass and charge. The detailed phase diagrams of $\epsilon_m$ and $\epsilon_q$ are shown in figures \ref{fig:emq_diagram1} and \ref{fig:emq_diagram2}. Since equation \eqref{eq:ratio} is symmetric under $(\epsilon_m, \epsilon_q) \rightarrow (1-\epsilon_m, 1-\epsilon_q)$, the figures are symmetric under the $180^\circ$ rotation at $(1/2,1/2)$. The initial charges are fixed as $Q=0.5$ and $Q=1.0$ in figures \ref{fig:emq_diagram1} and \ref{fig:emq_diagram2}, respectively. The chosen initial masses in figure \ref{fig:emq_diagram1} (a), (b), (c), (d) are $M=M_{\text{ext}}=0.544331$, $M=1.0$, $M=1.5$, $M=2.0$, and those in figure \ref{fig:emq_diagram2} (a), (b), (c), (d) are $M=M_{\text{ext}}=1.23132$, $M=1.5$, $M=2.0$, $M=2.5$, respectively.

Once again, the allowed region of charged AdS black hole has a bounded value $Q_{\text{ext}}$ for a given mass $M$. All the fragmented black holes as well as the initial one should belong to the allowed region. This puts constraint on $\epsilon_m$ and $\epsilon_q$. The black solid lines in figures \ref{fig:emq_diagram1} and \ref{fig:emq_diagram2} represent the boundary imposed by these constraints. We labelled a stable and unstable region in each figure and plot the boundary of instability as a red dashed line. Thin gray diagonal lines represent the uniform fragmentation ratios $\epsilon_m=\epsilon_q$, analyzed in the previous section. The red, blue and green markers with different shape along those lines represent the parameter choices $\epsilon_m=\epsilon_q=1/2$, $1/4$ and $1/8$, respectively.

The allowed regions of figures \ref{fig:emq_diagram1} and \ref{fig:emq_diagram2} covered by black solid line expand as the mass of initial black hole with a fixed charge increases. On the contrary, the allowed regions shrink when the charge increases with a fixed mass. This can be seen by comparing two figures \ref{fig:emq_diagram1} (c) and \ref{fig:emq_diagram2} (b) or figures \ref{fig:emq_diagram1} (d) and \ref{fig:emq_diagram2} (c). It means that the area of allowed region depends on the ratio of mass and charge of the initial black hole.

The behavior of instability boundary represented red dashed line is more complicated to understand. The allowed region consists of stable and unstable regions. At $M=M_{\text{ext}}$, most regions are unstable except the narrow stable regions close to the black solid line as can be seen in figure \ref{fig:emq_diagram1} (a). As mass increases beyond $M_{\text{ext}}$, the stable region grows and most of allowed regions are stable as shown in figure \ref{fig:emq_diagram1} (b). If the mass increases further, the stable region starts shrinking again as can be seen in figure \ref{fig:emq_diagram1} (c) and (d). This trends also can be seen in figure \ref{fig:emq_diagram2}. As mass increases starting with the mass $M_{\text{ext}}$, the unstable region first decreases as in figure \ref{fig:emq_diagram2} (b) and increases again in figure \ref{fig:emq_diagram2} (c) and (d). Now, figure \ref{fig:emq_diagram1} (c) with \ref{fig:emq_diagram2} (b) and figure \ref{fig:emq_diagram1} (d) with \ref{fig:emq_diagram2} (c) are compared for the fixed charge comparison. Then, the unstable region expands to the diagonal direction which implies the result of figure \ref{fig:mq_diagram}, but also the stable region moves to small $\epsilon_q$ area, because the increased charge makes more possibilities to be stable for small ratio of charge.

We now compare figure \ref{fig:emq_diagram1} with figure \ref{fig:mq_diagram}. Figure \ref{fig:emq_diagram1} corresponds to the straight horizontal line with charge $Q=0.5$ in figure \ref{fig:mq_diagram}. More specifically, figures \ref{fig:emq_diagram1} (a), (b), (c), and (d) corresponds to the markers 3(a), 3(b), 3(c), and 3(d) on that line in figure \ref{fig:mq_diagram} with the mass $M_{\text{ext}}$, $1.0$, $1.5$, and $2.0$, respectively. The fragmentation ratio value $\epsilon$ determines the position along the gray diagonal lines in figure \ref{fig:emq_diagram1}. For example, $\epsilon=1/2$ is represented by the red markers located at the center of each plots in figure \ref{fig:emq_diagram1}. The red markers in figures \ref{fig:emq_diagram1} (a), (b), (c), and (d) lies in the unstable, stable, unstable, and unstable regions, respectively. As can be seen in figure \ref{fig:mq_diagram} (a), the points 3(a), 3(b), 3(c), and 3(d) lies in the unstable, stable, unstable, and unstable regions, respectively, under fragmentation with $\epsilon=1/2$ represented by red dashed line. This exactly matches as represented with fixed $\epsilon=1/2$ in figure \ref{fig:emq_diagram1}. Similarly one can compare other values of fragmentation ratio in figure \ref{fig:mq_diagram} and figure \ref{fig:emq_diagram1} represented in different colors.

One can imagine the three-dimensional phase diagram of instability for the black hole fragmentation with respect to mass, charge of an initial black hole and uniform fragmentation ratio, $\epsilon_m=\epsilon_q$. Then, figure \ref{fig:mq_diagram} is mass versus charge section with a fixed fragmentation ratio and the diagonal lines of figures \ref{fig:emq_diagram1} and \ref{fig:emq_diagram2} are mass versus fragmentation ratio section with a fixed charge. 

\begin{figure}[ht]
\centering
\subfigure[(a)][$\ell=0.9$]{\includegraphics[width=0.32\textwidth]{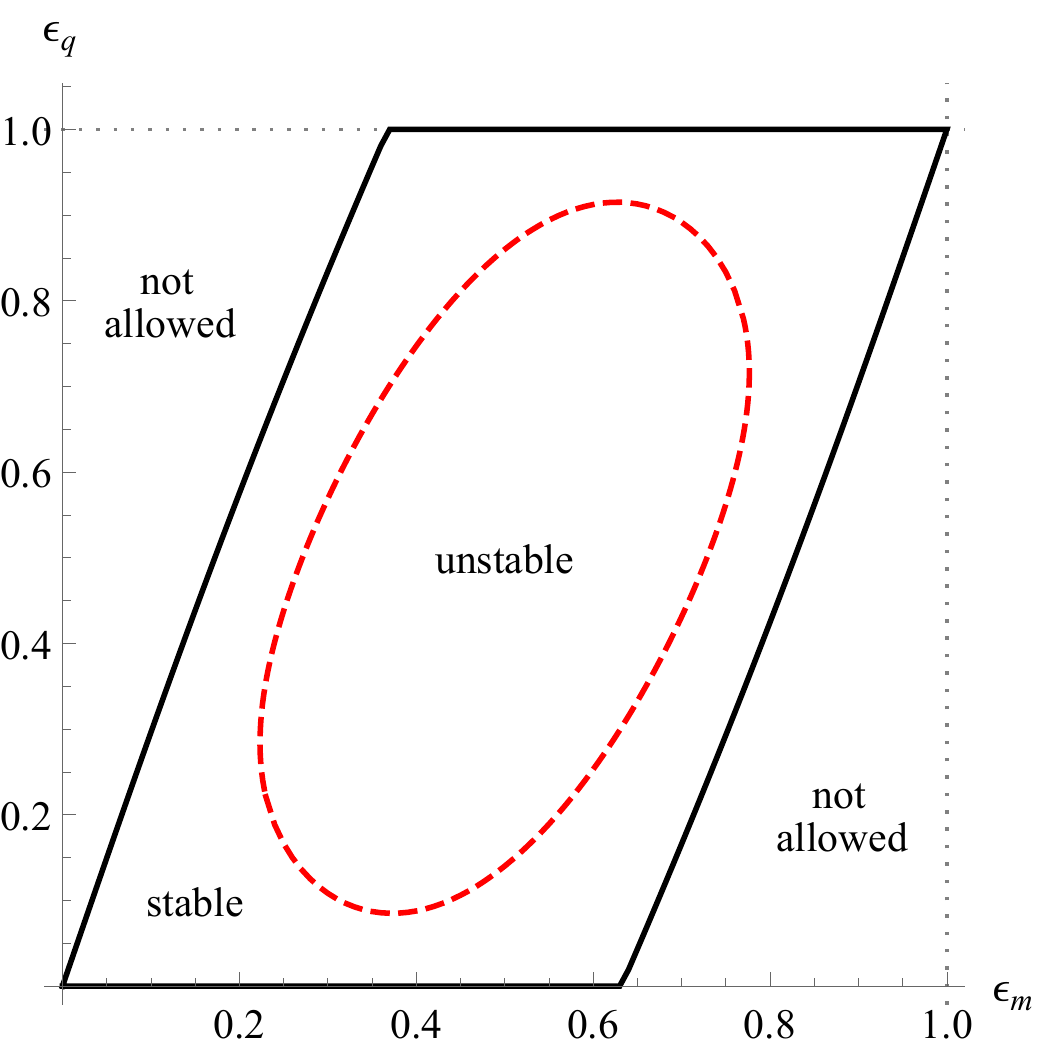}}
\subfigure[(b)][$\ell=1.0$]{\includegraphics[width=0.32\textwidth]{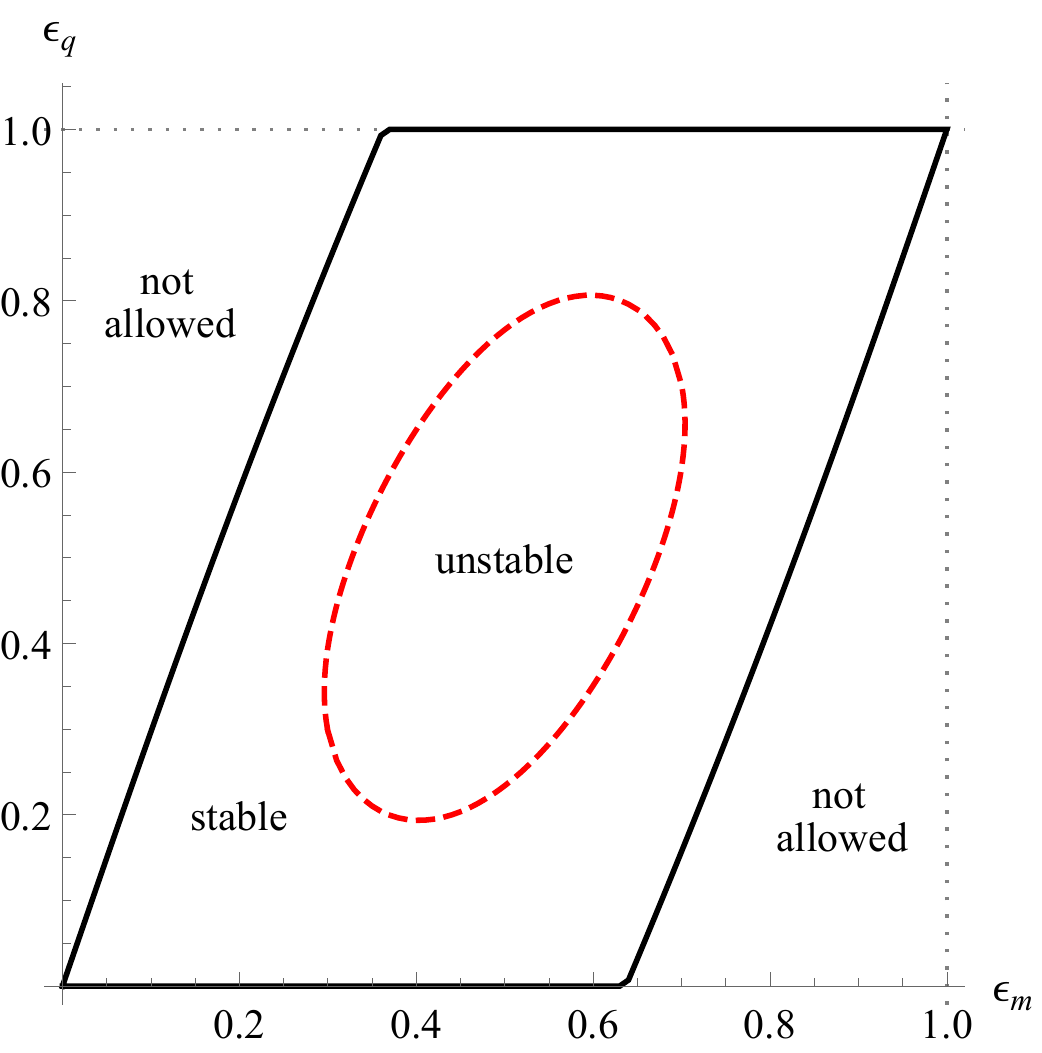}}
\subfigure[(c)][$\ell=1.1$]{\includegraphics[width=0.32\textwidth]{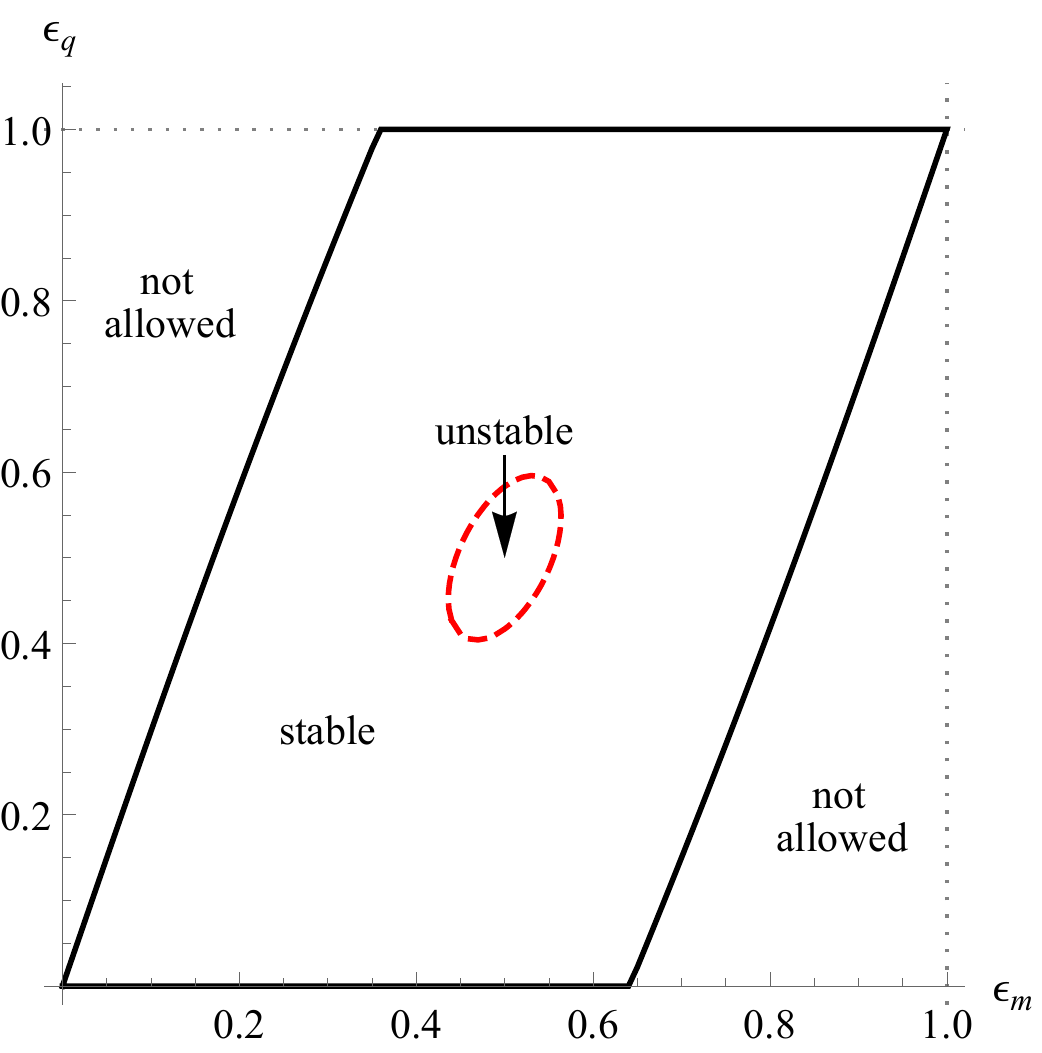}}
\caption{The phase diagrams for figure \ref{fig:emq_diagram1} (c) with different values of $\ell$.}
\label{fig:flat_limit}
\end{figure}

We take the flat space limit, $\ell \rightarrow \infty$ same as in section \ref{sec:3.2}. In that case, the variables are dimensionless and the flat space limit is equal to the small mass and charge limit. However, the mass and charge of initial black hole are already fixed in this section. Therefore, the original dimensionful parameters is used. The parameter values of figure \ref{fig:emq_diagram1} (c) are chosen for the flat space limit. In this limit, it makes unstable region contract to $(1/2,1/2)$. The behavior of shrinking unstable region implies that the black hole fragmentation only into two identical black hole is highly preferred as we discussed in previous paragraph. In addition, it is naturally expected that the flat space limit eliminates the unstable region for all parameter choices. Finally, the charged AdS black hole will be stable under the flat space limit even in non-uniform fragmentation. In other words, RN black hole should be stable under fragmentation with arbitrary ratio of mass and charge. This is consistent with known result that RN black hole is stable under the perturbation.

\section{Summary and Discussion} \label{sec:4}
We investigated the instability of charged AdS black holes with dimensions $D\geq 4$, under fragmentation. We assume a large AdS radius limit, large distance between fragmented black holes to ignore the gravitational and electromagnetic forces, and $\ell \gg R$. Under these conditions, the mass is conserved in the fragmentation process. The black hole with a mass $M$ and charge $Q$ is fragmented into two black holes one of which has the mass $\epsilon_m M$ and charge $\epsilon_q Q$ and the other the mass $(1-\epsilon_m) M$ and charge $(1-\epsilon_q) Q$. The instability of black holes is obtained by comparing the entropy of initial with that of final black hole states. 

Instability of charged AdS black holes is expected from the analytical calculation of entropy ratio in the limit of small and large masses as shown in table \ref{tab:analytic}. The charged AdS black hole is stable in the small mass limit while unstable in the large mass limit. Thus, the instability boundary should exist in between. We first investigated the case of uniform fragmentation, $\epsilon_m=\epsilon_q$. The instability with respect to the mass and charge of black hole is shown in figure \ref{fig:mq_diagram}. The black hole becomes unstable, stable and again unstable as mass increases with fixed charge, but it becomes unstable as charge increases with fixed mass. There exists an $\epsilon$ which makes the black hole unstable with minimal charge for each masses. It become zero when the mass goes to zero while it increases toward and become $1/2$ as mass increases until the maximum mass of instability boundary with $\epsilon=1/2$. For higher-dimensions, the phase structure is qualitatively same except that stable regions are shrinking.

In figure \ref{fig:mem_diagram}, we represents the boundary mass of extremally charged AdS black hole $M_{\text{ext},\epsilon}$ and Schwarzschild-AdS black hole $M_{\text{crit},\epsilon}$ in order to show the intersection of boundaries. Indeed, there is an intersection due to the contradictory trends of figures \ref{fig:mem_diagram} (a) and (b). Note that, the critical mass $M_{\text{crit},\epsilon}$ in figure \ref{fig:mem_diagram} (b) goes to infinity in the small ratio limit, $\epsilon \rightarrow 0$. It means Schwarzchild-AdS black hole is stable under fragmentation with a small ratio value, which can be treated as a perturbation. This is consistent with the fact that Schwarzchild-AdS black hole is stable under the perturbation.

For general fragmentation case, the instability of charged AdS black holes with respect to the mass and charge ratio, $\epsilon_m$ and $\epsilon_q$, is shown in figures \ref{fig:emq_diagram1} and \ref{fig:emq_diagram2}. The allowed region increases as mass increases, but it decreases as charge increases. The stable region first increases as mass increases and decreases again as mass further increases for fixed charge. In addition, it always decreases as charge increases for fixed mass. There are labels in figure \ref{fig:mq_diagram} that matches the parameter values of labelled figures. Furthermore, the color of markers in figures \ref{fig:emq_diagram1} and \ref{fig:emq_diagram2} indicate the value of uniform fragmentation ratio, $\epsilon$. One can think the figures as a section of the full three-dimensional phase diagram composed of the mass, charge of black hole and uniform fragmentation ratio.

We studied the instability of charged AdS black holes by comparing the entropy ratio of initial and final black hole states. However, we have not studied the dynamical or tunnelling process that leads to the instability. The entropy ratio smaller than 1 is sufficient to guaranty the stability. On the contrary, the entropy ratio larger than 1 is not sufficient but necessary condition for the black hole to be unstable under fragmentation. This is because the lifetime of initial state can be either short or very long depending on the fragmentation process. The process leading to the fragmentation may occur dynamically or through tunnelling due to the thermal or quantum fluctuation. Investigation on this aspect is left for future work.

\section*{Acknowledgements}
This work was supported by the National Research Foundation of Korea(NRF) grant funded by the Korea government(MSIP)(No.~2014R1A2A1A010). BG was supported by Basic Science Research Program through the National Research Foundation of Korea(NRF) funded by the Ministry of Science, ICT \& Future Planning(2015R1C1A1A02037523). BG would like thank for hospitality to the members of APCTP where part of this work was done during visiting.

\bibliographystyle{jhep}
\bibliography{references}

\end{document}